\documentclass[twocolumn,secnumarabic,amssymb, nobibnotes, aps, prl,superscriptaddress,floatfix]{revtex4-2}

\usepackage{graphicx}
\usepackage{gensymb}
\usepackage{wrapfig}
\usepackage{amsmath}
\usepackage{xcolor}
\usepackage{physics}
\usepackage{siunitx}
\usepackage{upgreek}
\usepackage{braket}
\usepackage[colorlinks,linkcolor=black,citecolor=black,urlcolor=black,filecolor=black]{hyperref}
\begin{document}

\title{Reporter-spin-assisted \(T_1\) relaxometry}

\author{Zhiran Zhang} 
\affiliation{Department of Physics, University of California, Santa Barbara, California 93106, USA}
\author{Maxime Joos}
\affiliation{Department of Physics, University of California, Santa Barbara, California 93106, USA}
\author{Dolev Bluvstein}
\affiliation{Department of Physics, University of California, Santa Barbara, California 93106, USA}
\affiliation{Department of Physics, Harvard University, Cambridge, Massachusetts 02138, USA}
\author{Yuanqi Lyu}
\affiliation{Department of Physics, University of California, Santa Barbara, California 93106, USA}
\affiliation{Department of Physics, University of California, Berkeley, California 94720, USA}
\author{Ania C. Bleszynski Jayich}
\affiliation{Department of Physics, University of California, Santa Barbara, California 93106, USA}

\email[email:]{ania@physics.ucsb.edu}

\begin{abstract}
A single spin quantum sensor can quantitatively detect and image fluctuating electromagnetic fields via their effect on the sensor spin's relaxation time, thus revealing important information about the target solid-state or molecular structures. However, the sensitivity and spatial resolution of spin relaxometry are often limited by the distance between the sensor and target. Here, we propose an alternative approach that leverages an auxiliary reporter spin in conjunction with a single spin sensor, a diamond nitrogen vacancy (NV) center. We show that this approach can realize a $10^4$ measurement speed improvement for realistic working conditions and we experimentally verify the proposed method using a single shallow NV center. Our work opens up a broad path of inquiry into a range of possible spin systems that can serve as relaxation sensors without the need for optical initialization and readout capabilities.
\end{abstract}

\maketitle

The detection of fluctuating electromagnetic fields lends important insight into the dynamics of solid-state systems, for example, the local current and spin fluctuations in magnetic and correlated electron systems \cite{PhysRevB.20.850,Agarwal2017,Flebus2018,Khoo2021,Chatterjee2022,Dolgirev2022}, decoherence processes in quantum systems \cite{Schoelkopf2003,Rosskopf2014,Romach2015,Myers2014, Myers2017}, and chemical and biological processes \cite{Caravan1999,Raitsimring2007,Giannoulis2021,Li2022}. 
Single-spin quantum sensors constitute a powerful tool for detecting fluctuating fields; in a technique called relaxometry, fluctuating fields with a spectral component matched to the energy splitting of the sensor spin reduce the spin's relaxation time $T_1$ \cite{Degen2017}. Single-spin relaxometry features noninvasive and quantitative measurement of the fluctuating fields as well as high spatial resolution down to the nanometer scale. %


Nitrogen vacancy (NV) centers in diamond are a prominent example of a solid-state spin qubit sensor, exhibiting a wide temperature operating range, compatibility with other systems, high sensitivity, and high spatial resolution.
Relaxometry with NV centers has been used to probe magnetic fluctuations near the diamond surface to better understand surface-induced decoherence \cite{Myers2014,Rosskopf2014,Romach2015,Myers2017}, detect spin waves in magnetic systems \cite{VanDerSar2015,Du2017}, image local conductivity and current flow of condensed matter systems \cite{Kolkowitz2015,Ariyaratne2018,Andersen2019}, perform spectroscopy of electronic spins \cite{Hall2016}, and detect magnetic nanoparticles \cite{Schmid-Lorch2015} and magnetic ions \cite{Steinert2013,Pelliccione2014,Sushkov2014a}.
The proximity of the sensor to its target is critical to achieving high spatial resolution and high sensitivity, and becomes particularly important for relaxometry when targeting the detection of single spins (nuclear or electronic), as the dipolar magnetic fluctuation signal from a single spin die off as $1/r^6$, with $r$ being the sensor-target separation \cite{Steinert2013,Tetienne2013a}. Further, the need for proximity is made more acute in imaging experiments when long measurement times can lead to significant drifts in the sensor-target distance that may render the images unrecognizable. 


Bringing NV centers close to the diamond surface is one natural approach to reduce sensor-target separation for improved relaxometry. However, NV centers with high-grade properties cannot be made arbitrarily shallow for many reasons: firstly, the yield rate of an implanted nitrogen atom forming an NV center declines dramatically near the surface  \cite{Pezzagna2010}, and secondly, near-surface NVs tend to exhibit increased charge instabilities \cite{Rondin2010,Bluvstein2019,Yuan2020} and shorter coherence times \cite{Myers2014,Sangtawesin2019}. Overcoming these challenges is an active area of study.


\begin{figure}
\includegraphics[width=86mm]{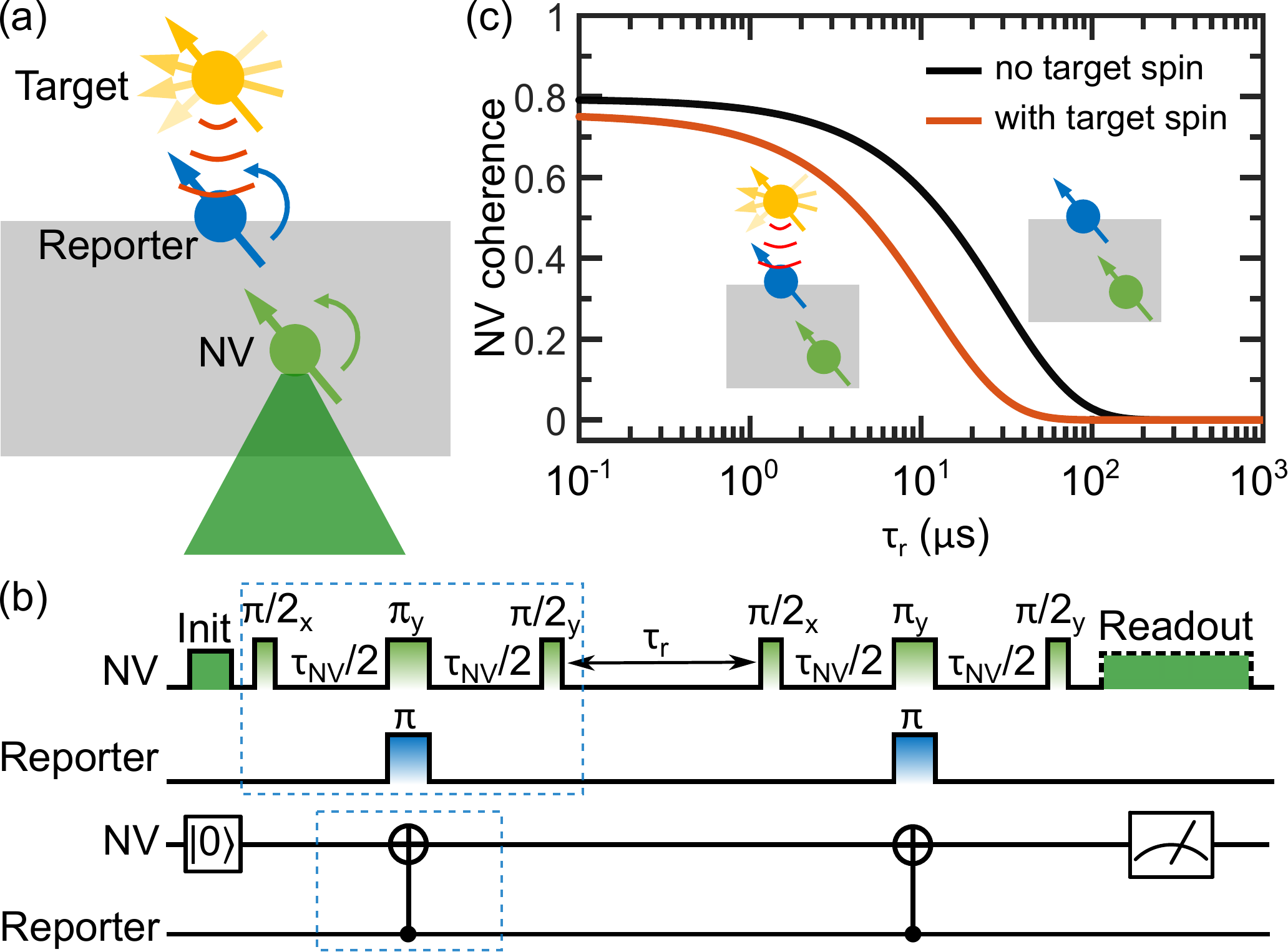}
\caption{\label{fig1}(a) Schematic of the proposed experiment. To detect the magnetic fluctuations (red contours) from a target spin (yellow), an optically addressable NV center in diamond senses a change in the relaxation time $T_{1,R}$ of a reporter spin at the diamond surface. The close proximity of the reporter spin to the target spin amplifies the signal. (b) Pulse sequence and corresponding quantum circuit diagram (bottom) \cite{Rezai2022} used for measuring $T_{1,R}$. After optical initialization of the NV center (dark green), microwave pulses control the spin states of NV center (light green) and reporter spin (blue), followed by optical readout of the NV (dark green).(c) Simulated NV coherence as a function of $\tau_r$ as measured by the pulse sequence shown in (b), for an NV 4.5 nm deep. In the absence of a target spin, the black curve shows the signal corresponding to the reporter spin's intrinsic $T_1$, 30 $\mu$s in this case. The red curve shows a faster decay when a nearby Gd$^{3+}$ spin, 3 nm from the reporter spin in this case, reduces $T_{1,R}$ to \SI{11.6}{\micro s}.}
\end{figure}

Here we propose an alternative approach that leverages an auxiliary spin that resides closer to, or even at, the diamond surface (Fig. \ref{fig1}(a)) to sense fluctuating fields. This reporter spin acts as the relaxation sensor, whereas a nearby NV center, comfortably deeper in the diamond, serves as a local optical readout of the reporter spin state \cite{Schaffry2011,Sushkov2014}. Compared to direct NV relaxometry, this method features improved sensitivity and spatial resolution while circumventing the reduced NV coherence and charge stability associated with the diamond surface. In essence, the main advantage of the reporter relaxometry method stems from the fact that the reporter translates an incoherent magnetic field signal, which decays as $1/r^6$, into a coherent magnetic signal emanating from the reporter spin with a $1/r^3$ dependence. Furthermore, the proposed approach offers access to an additional range of detection frequencies determined by the reporter spin's energy splitting, which is distinct from the NV sensor's splitting. In this paper, we analytically examine the dependence of the relaxation signal on NV and reporter spin properties, finding a measurement speed increase up to $10^4$-fold compared to conventional NV relaxometry as relevant parameters are varied in real working conditions. For concreteness, we benchmark performance using a specific example of detecting and imaging a single Gadolinium (Gd$^{3+}$) ion, a commonly used spin label for bio-structural imaging, but we remark that the results are broadly applicable to other target systems. We then experimentally verify the proposed pulse sequence with a single NV center strongly coupled to a nearby reporter spin, whose relaxation time is tuned via a stochastic driving technique \cite{Joos2021}. Finally, the challenges and future outlook of this novel approach are discussed.

We consider a single reporter spin located at the diamond surface near a single NV center, as shown in Fig. \ref{fig1}(a). Although the reporter spin can come in any form, its primary requirement is a long intrinsic $T_1$. We note that single spins at the diamond surface have been detected with 100-$\mu$s-scale relaxation times \cite{Sushkov2014a,DeWit2018}, which are sufficiently long for the protocols proposed here. For simplicity, we discuss the case of spin-1/2 reporter spins, but the analysis can be extended to systems with larger spins. 

The proposed reporter-spin relaxometry protocol is shown in Fig.~\ref{fig1}(b). This protocol probes the correlation time of the magnetic field signal produced by the reporter spin, which is equal to its relaxation time $T_{1,R}$, via its dipolar coupling to the NV using double electron electron resonance (DEER) techniques. The sequence constitutes a correlation measurement of the NV center's environment seen through the filter function set by the NV pulse sequence, an ``xyy" Hahn echo like sequence in this case. Importantly, by matching $\tau_\text{NV}$, to the inverse of the NV-reporter dipolar spin coupling rate $k_s$, the sequence selectively probes the coupling between the reporter spin and NV. Therefore, the two separate ``xyy" DEER sequences are equivalent to two CNOT gates in quantum circuit representation \cite{Rezai2022}. The correlation time of such coupling is then imprinted onto the NV coherence, which can be measured via differential photoluminescence readout of the NV center's spin state ~\cite{Myers2017}. In effect, the NV center acts as a ``flag'' qubit whose state changes if the reporter spin flips during the correlation sequence \cite{PhysRevLett.121.050502}. We note that to probe more weakly coupled reporter spins, one has to extend $\tau_\text{NV}$, and the Hahn echo may need to be replaced by dynamical decoupling sequences such as XY8 with corresponding microwave pulses on the reporter spin.


Figure~\ref{fig1}(c) shows the expected signal for the example case of detecting a single, proximal fluctuating Gd$^{3+}$ spin, a spin label with a large electronic spin of $S=7/2$ and fast GHz-scale spin dynamics \cite{Caravan1999,Goldfarb2014}; ensembles of Gd$^{3+}$ spins have been interfaced with and detected by NV centers  \cite{Steinert2013,Tetienne2013a,Sushkov2014a,Pelliccione2016}. The Gd$^{3+}$ produces a rapidly fluctuating magnetic field, which 
reduces the correlation time of the reporter spin and manifests clearly as a faster NV population decay. (See SI for details.) The small reduction in NV coherence as can be seen from short $\tau_r$ is caused by the relaxation of reporter spin during the ``xyy" DEER. The NV parameters used in the simulations are experimentally measured on an implanted shallow NV (NV1) in a chemical vapor deposition-grown diamond sample. The parameters are $T_2=\SI{8.4}{\micro s}$, and $T_{1,\text{NV}}=\SI{3.5}{ms}$,  and the NV depth is measured via proton NMR \cite{Pham2016,Bluvstein2019a} to be \SI{4.5}{\nano m}.
The reporter spin is assumed to be located on the diamond surface at a position where the dipolar coupling to the NV is maximized, $T_{1,R}$ is assumed to be \SI{30}{\micro s}, and $\tau_\text{NV}$ is set to 912 ns to match the inverse of the dipolar coupling strength $k_s$ \cite{Sushkov2014}.

To quantitatively compare the performance of the proposed reporter-spin-assisted relaxometry protocol with direct NV relaxometry, we first discuss how a target fluctuating magnetic field external to the diamond imprints itself on the relaxation time of a single spin (either the NV center or the reporter spin):
\begin{equation}
    \frac{1}{T_{1}} =\frac{1}{T^{'}_{1}}+N_S\frac{\gamma_\text{NV}\gamma_R}{2}\left[S_{B_x}(\omega)+S_{B_y}(\omega)\right], \label{equation1}
\end{equation}
where $T^{'}_{1}$ is the spin's intrinsic relaxation time without the external fluctuating fields, $\gamma_\text{NV}$ and $\gamma_R$ are the gyromagnetic ratios of NV and reporter spin, $S_{B}$ is the noise spectral density of the magnetic field experienced by the spin, $N_S=3$ for the NV spin (or $N_S=2$ for spin-1/2 reporter), and $\omega$ is the transition frequency of the spin. 
Assuming a Lorentzian spectrum of the fluctuating field, Eq.~(\ref{equation1}) can be written as
\begin{equation}
\begin{split}
    \frac{1}{T_1}&=\frac{1}{T_1 ^{'}}+N_S\gamma_\text{NV}\gamma_{R}\langle B_\perp^2\rangle\frac{\tau_c}{1+\omega ^2\tau_c^2},
\end{split}
\label{equation2}
\end{equation}
where $\langle B_\perp^2 \rangle=\langle B_x^2\rangle+\langle B_y^2\rangle$ is the variance of the magnetic field transverse to the quantization axis of the spin and is proportional to $1/r^6$ (see Supplementary Material for details of derivation), and $\tau_c$ is the correlation time of magnetic field from a fluctuating Gd$^{3+}$, which we take to be 0.35 ns, as reported in the literature \cite{Kim2009}.

\begin{figure}
\includegraphics[width=86mm]{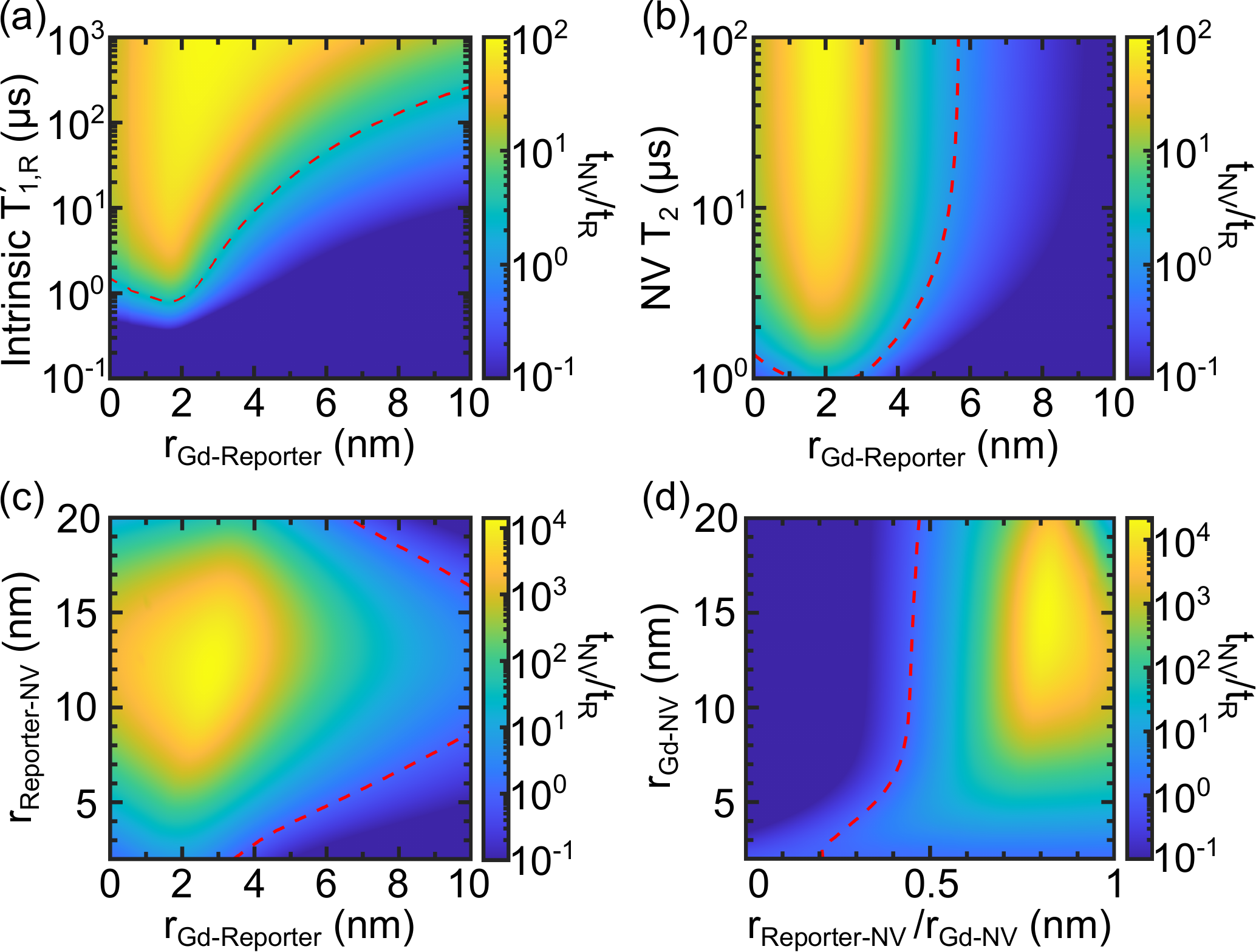}
\caption{\label{fig2}Speed enhancement of reporter relaxometry over direct NV relaxometry for single Gd$^{3+}$ spin detection. (a) Speed enhancement ($t_{\text{NV}}$/$t_{R}$) as a function of the distance from the Gd$^{3+}$ to reporter ($r_\text{Gd-Reporter}$) and the intrinsic $T_1$ of the reporter. We assume NV $T_2=\SI{8.4}{\micro s}$, and NV depth is 4.5 nm. (b) Speed enhancement ($t_{\text{NV}}$/$t_R$) as a function of the distance from the single Gd$^{3+}$ to the reporter, and the NV $T_2$. We assume reporter spin $T_{1,{R}}=\SI{30}{\micro s}$, and NV depth is 4.5 nm. (c)(d) Speed enhancement ($t_{\text{NV}}$/$t_R$) as a function of the distances between NV, reporter, and the single Gd$^{3+}$. We assume NV $T_2=\SI{100}{\micro s}$ and reporter spin $T_{1,{R}}=\SI{30}{\micro s}$. SCC readout technique is used here for both reporter relaxometry and NV relaxometry. The red dashed lines indicate a speed enhancement of 1.
}
\end{figure}

Figure \ref{fig2} plots the speed enhancement of the reporter spin relaxometry protocol over the direct NV relaxometry protocol, varying several parameters to highlight in which situations reporter spins are an advantageous choice. The qualitative picture that emerges from the four plots is that longer intrinsic reporter $T_{1,R}^{'}$, longer NV $T_2$, smaller reporter-Gd$^{3+}$ separations, and deeper NV centers enhance the benefits of reporter relaxometry, culminating in a 10$^4$-fold speed enhancement for a 10-15 nm deep NV with $T_\text{2}=\SI{100}{\micro s}$ and a Gd$^{3+}$ spin located $\sim$3 nm above a reporter spin with $T_{1,R}^{'}=\SI{30}{\micro s}$ (Fig.~\ref{fig2}c-d). We note that these are all experimentally confirmed values \cite{Sushkov2014a,Myers2017,Bluvstein2019,Bluvstein2019a}.  A lower (higher) NV $T_2$ would shift the location of maximal speed enhancement in Fig.~\ref{fig2}c to smaller (larger) reporter-NV separations ($r_\text{reporter-NV}$), and reduce (enhance) the speed enhancement value ($t_\text{NV}/t_R$) (plots are shown in the SI). We assume the use of the spin-to-charge conversion (SCC) readout technique for all cases here, where the readout noise level is experimentally verified on NV1 \cite{Shields2015}.  
The speed enhancement is obtained by computing the ratio $t_{NV}/t_{R}$; $t_{R}$ is the averaged minimal time required to detect a reduction in reporter $T_{1,R}$ if using reporter relaxometry,
\begin{equation}
    t_{R}=\frac{(\text{SNR})^2C_{\text{SPN}}^2}{2(\Delta S)^2}t_{\text{seq}},
\label{eqn:tr}
\end{equation}

where $\Delta S$ is the change of the signal due to the reduced $T_{1,R}$, $\text{SNR}$ is the desired signal-to-noise ratio, $C_\text{SPN}$ is the ratio between experimental measurement uncertainty and the spin projection noise limit, and $t_\text{seq}$ is the total duration of the pulse sequence including the initialization and readout time. $t_\text{NV}$ is computed analogously using Eq.~\ref{eqn:tr} with the corresponding $\Delta S$ and $t_\text{seq}$ for direct NV relaxometry. For each point in the simulations in Fig.~\ref{fig2}, the readout and measurement times are optimized to minimize $t_{R}$ and $t_\text{NV}$ separately, and the details are discussed in SI. We note that the speed enhancements shown in Fig.~\ref{fig2} will be even more significant if a standard 532 nm NV photoluminescence readout is used instead of SCC readout techniques, because SCC readout techniques are more effective for the longer pulse sequences associated with direct NV relaxometry compared to reporter relaxometry\cite{Ariyaratne2018}.

\begin{figure}
\includegraphics[width=86mm]{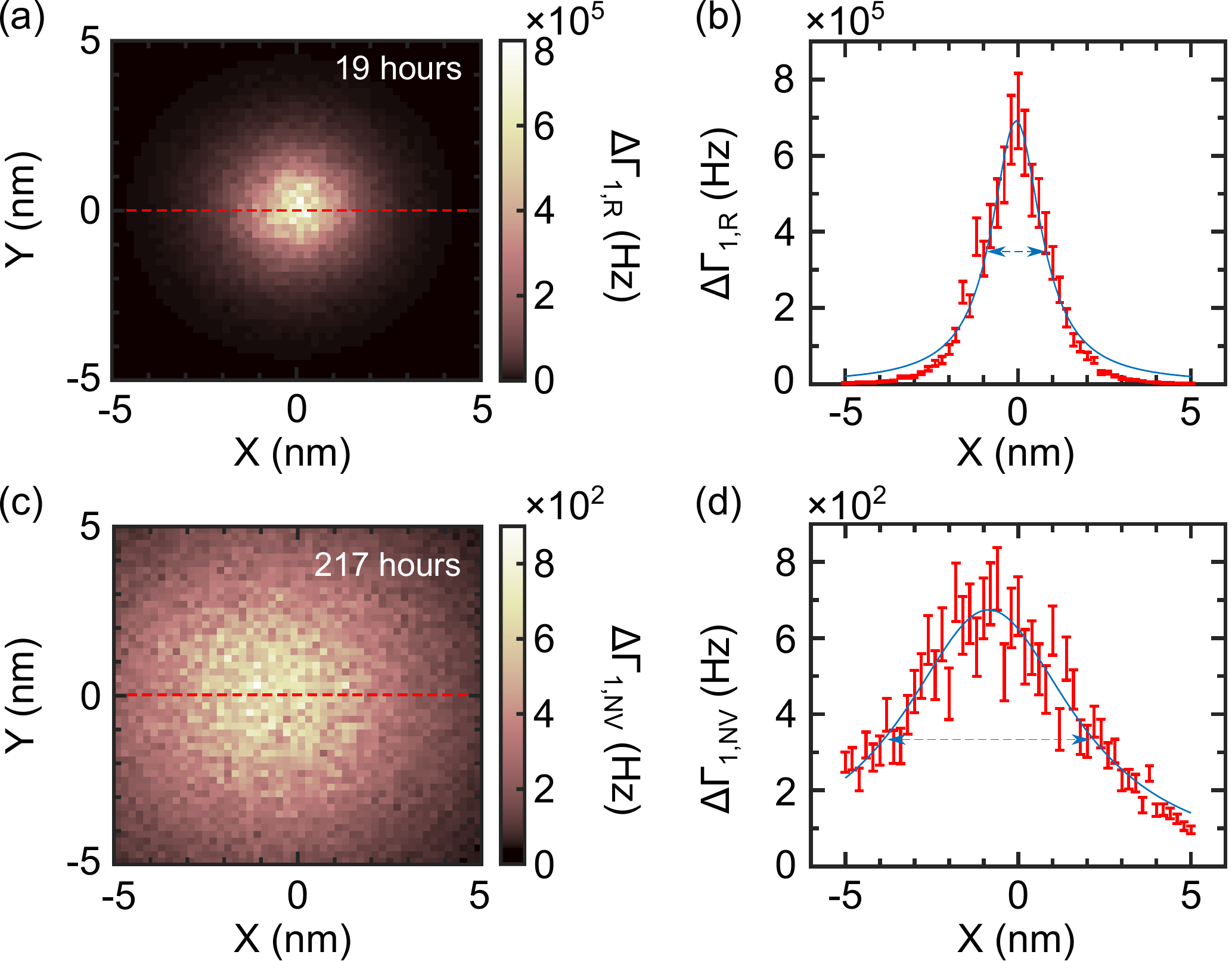}
\caption{\label{fig3}A comparison of simulated scanning images of a single Gd$^{3+}$ spin acquired by reporter relaxometry and direct NV relaxometry. (a) Reporter-spin assisted relaxometry image: plotted is the change of reporter relaxation rate, $\Delta\Gamma_{1,{R}}$, as the reporter spin is scanned in a plane 2 nm above a single Gd$^{3+}$. (b) A line-cut along the red dashed line in (a). (c) Direct NV relaxometry image plots the change of NV relaxation rate, $\Delta\Gamma_{1,\text{NV}}$, as the NV center is scanned in a plane 6.5 nm above a single Gd$^{3+}$. (d) A line-cut of the red dashed line in (c). The reporter relaxometry image in (a) takes 19 hours compared to 217 hours for direct NV relaxometry imaging. These simulations target the same level of relative standard error for each pixel. The solid blue lines in (b) and (d) are Lorentzian fits, and the dashed lines with arrows indicate the FWHM. 
For both images, the NV $T_{1,\text{NV}}=\SI{3.5}{ms}$, $T_2=\SI{8.4}{\micro s}$, and NV depth is 4.5 nm. The reporter $T_{1,{R}}=\SI{100}{\micro s}$ and is located on the diamond surface. 
}
\end{figure}

Reporter relaxometry can also be combined with scanning probe microscopy (SPM) \cite{Pelliccione2014,Schmid-Lorch2015,Ariyaratne2018,Finco2021} to achieve better spatial resolution than conventional NV relaxometry imaging in a given measurement time as well as providing faster imaging for a given sensitivity. In reporter-spin-assisted scanning relaxometry, a reporter spin is incorporated onto the apex of a diamond scanning probe tip with a nearby subsurface NV center and is scanned over an imaging target, spatially mapping the fluctuating fields emanating from the sample. 
In Fig.~\ref{fig3} we compare two simulated images of a single Gd$^{3+}$ spin obtained using scanning reporter $T_1$ relaxometry (Fig. \ref{fig3}a) and direct scanning NV $T_1$ relaxometry (Fig. \ref{fig3}b). The change of the reporter's relaxation rate is plotted against its lateral position relative to the Gd$^{3+}$ as it is scanned above the diamond surface. We use an adaptive measurement technique \cite{Ariyaratne2018} for both relaxometry methods and set the averaging time at each pixel to maintain a constant relative standard error of $\Delta\Gamma_1$, where $\Delta\Gamma_1=1/T_1-1/T_1^{'}$ is the change in relaxation induced by the Gd$^{3+}$. We find that reporter relaxometry shows a roughly 10-fold overall speed enhancement, 3.5-fold better spatial resolution, and produces a much more prominent signal as seen by comparing the signals shown in Fig.~\ref{fig3} (note the different color scale values). The center of the signal is offset from the actual Gd$^{3+}$ position in Fig. \ref{fig3} because the relaxation effect has an angular dependence on sensor spin's quantization axis and location of the Gd$^{3+}$. (See SI for details.) 

\begin{figure}
\includegraphics[width=86mm]{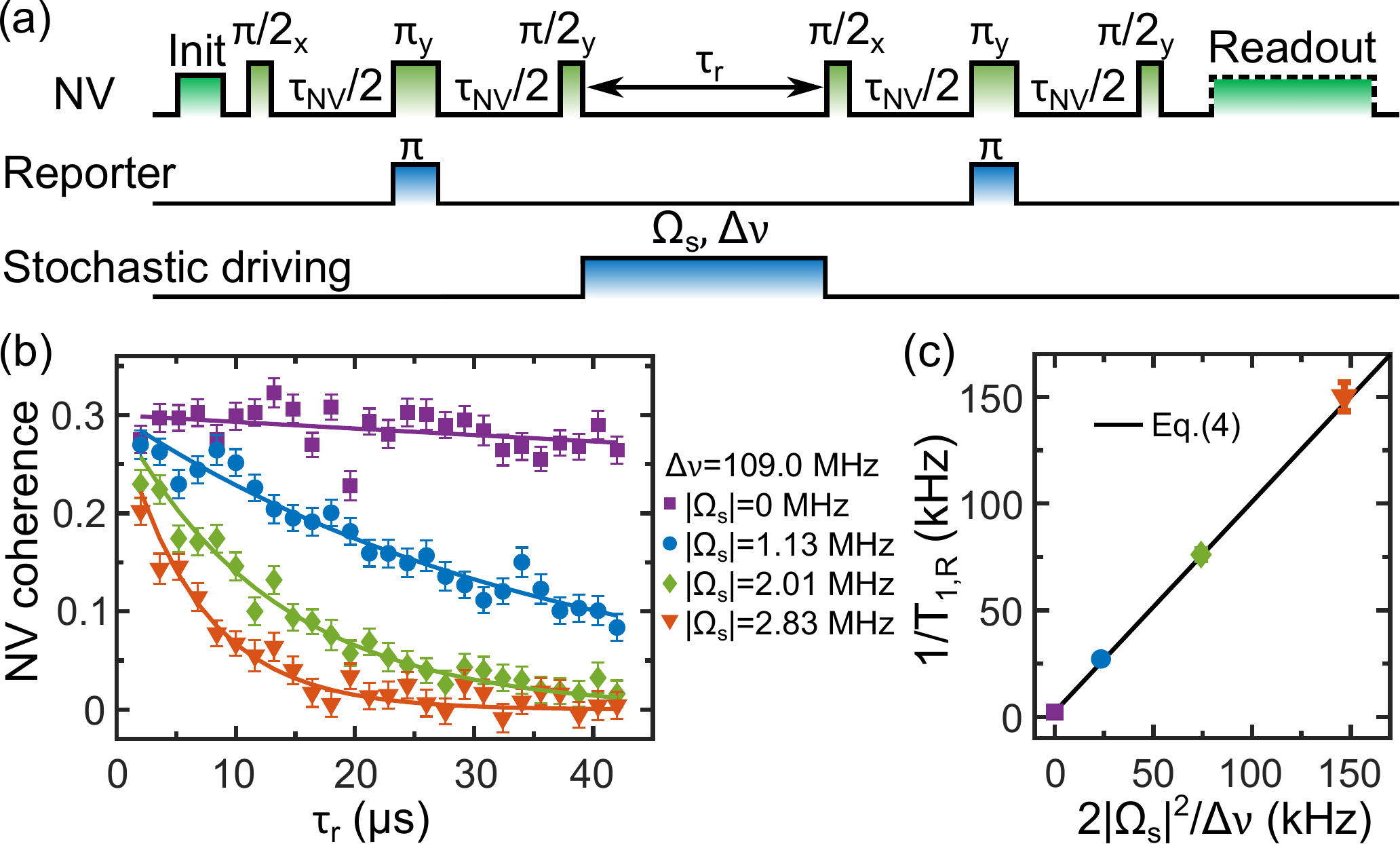}
\caption{\label{fig4}Demonstration of reporter relaxometry with artificially reduced reporter spin correlation by stochastic driving. The measurement is performed with a single NV center that is strongly coupled to a nearby g=2 spin-1/2.
(a) Pulse sequence of reporter relaxometry with additional stochastic driving of the reporter spin during $\tau_r$.
Incoherent spin dynamics caused by stochastic driving with Rabi frequency $\abs{\Omega_{s}}$ and linewidth $\Delta \nu$ reduces $T_{1,R}$ of the reporter spin.
(b) NV coherence for various stochastic driving powers, indicating reduced reporter spin auto-correlation with increased $\Omega_{s}$.
Solid lines are mono-exponential decay fits.
(c) Extracted reporter spin relaxation rate as a function of $2\abs{\Omega_{s}}^2/\Delta\nu$.
Black solid line is the theoretical behavior expected from Eq. (\ref{equation4}).
}
\end{figure}

We now experimentally demonstrate the ability of the proposed sequence (Fig. 1(b)) to accurately detect the $T_1$ reduction of a reporter spin. In this proof-of-principle experiment, we apply an external stochastic field \cite{Joos2021,Davis2021} to a reporter spin located in close proximity to an NV center, thus emulating the effect of fluctuating fields produced by a sensing target. The polychromatic drive, spectrally centered on the reporter spin resonance, reduces the correlation time of the reporter spin by inducing incoherent spin dynamics; the induced relaxation rate is controlled by the amplitude and broadening of the engineered field: 
\begin{equation}
\frac{1}{T_{1,R}}=\frac{1}{T^{'}_{1,R}}+2\frac{\abs{\Omega_{s}}^2}{\Delta \nu}.
\label{equation4}
\end{equation}
where $\abs{\Omega_{s}}$ is the Rabi frequency of the stochastic drive, and $\Delta \nu$ is the full-width at half maximum linewidth of the Lorentzian spectrum of the drive.
We implement the reporter-assisted relaxometry sequence (Fig. 4(a)) on a single shallow NV center (NV2) in diamond that is strongly coupled to a nearby g=2 reporter spin and we probe the correlation of the reporter spin while turning on stochastic driving centered at 888.0 MHz during $\tau_r$. (See SI for details.)
Figure~\ref{fig4}(b) shows the reduced correlation of the reporter spin mapped onto the NV coherence as the strength of the stochastic drive is increased.
For negligible stochastic drive power ($\abs{\Omega_{s}} = 0 \text{ kHz}$), the observed correlation is governed by the intrinsic slow relaxation of the reporter spin, $T'_{1,R} \approx \SI{1} {\milli s}$ in this case.
As the drive power is increased, the reporter spin's correlation time is reduced and dominated by its incoherently driven dynamics.
Figure~\ref{fig4}(c) shows the reporter spin decay rate extracted from a mono-exponential fit to the data in Fig.~\ref{fig4}(b) as a function of $\abs{\Omega_{s}}^2/\Delta\nu$.
Experimental results agree quantitatively with the expected behavior of Eq. (\ref{equation4}) demonstrating the suitability of the reporter spin-assisted relaxometry sequence.

Another benefit of reporter spin relaxometry is that it can probe fluctuating fields in a different frequency range than direct NV relaxometry, in particular giving access to a lower frequency range in moderate static magnetic fields (the NV probes higher frequencies because of its large zero-field splitting). 
Probing a lower frequency range provides a stronger fluctuating signal for many types of noise baths, such as a Lorentzian noise spectrum. We note that the speed and spatial resolution comparisons in Fig. \ref{fig2} and \ref{fig3} conservatively assume identical NV and reporter spin transition frequencies, but a smaller reporter spin transition frequency could yield a larger reduction in the relaxation time. Further, reporter spin relaxometry can be used in conjunction with NV relaxometry to gain more spectral information about the sensing target.
Our work opens up a broad path of inquiry into a range of possible reporter spin systems that can serve as relaxation sensors without the need for optical initialization and readout capabilities. 
While engineering single reporter spins at the diamond surface is challenging, there are several promising candidates. Naturally occurring surface spins located on the diamond surface have been detected and measured to have remarkably long $T_1=$ \SI{100}{\micro s} \cite{Grotz2011, Grinolds2014, Sushkov2014, Bluvstein2019a, Sangtawesin2019, Stacey2019,Dwyer2021}, though further work is necessary to confirm their microscopic origin and engineer their properties.
Reporter spins can also be engineered via ion implantation or chemical synthesis and patterning of molecules \cite{Morton2007,Bader2014}, ions encapsulated in fullerene \cite{Pinto2020}, rare-earth ions, and radical spin labels \cite{Kveder2008,Shi2018}.  

In conclusion, we propose a novel method that utilizes reporter spins in conjunction with optically addressable NV centers in diamond to improve the measurement sensitivity and spatial resolution of conventional NV $T_1$ relaxometry sensing and imaging.
We quantitatively compare the speed and spatial resolution of this method to conventional NV $T_1$ relaxometry, and find a wide range of parameter space in which reporter spin relaxometry provides substantial gains. Proof-of-principle experiments confirm the ability of the proposed sequence to quantitatively probe the relaxation of a single, dark reporter spin. This work motivates the development of engineered reporter spins and some candidates are proposed.


We thank Tian-Xing Zheng and Peter Maurer for helpful discussions. We gratefully acknowledge support from the US Department of Energy (BES grant No. DE-SC0019241) for surface spin studies, the DARPA DRINQS program (Agreement No. D18AC00014) for driving protocols, and the NSF Convergence Accelerator (Award No. 2040520) for relaxometry simulations. We acknowledge the use of shared facilities of the National Science Foundation (NSF) Materials Research Science and Engineering Center (MRSEC) at UC Santa Barbara, DMR 1720256, and the NSF Quantum Foundry through Q-AMASE-i
program award DMR-1906325. D.B. acknowledges support from the NSF Graduate Research Fellowship Program (grant DGE1745303) and The Fannie and John Hertz Foundation.
\bibliographystyle{apsrev4-2}
\bibliography{references}

\end{document}